\documentclass[12pt]{iopart}
\usepackage{graphicx}
\graphicspath{{figures/}}

\usepackage{iopams}
 
\begin{document}

\title{Regular Bardeen AdS Black Hole as a Heat Engine}

\author{Rajani K.V., Ahmed Rizwan C.L., Naveena Kumara A.,\\ Deepak Vaid \& Ajith K.M.}

\address{Department of Physics, National Institute of Technology Karnataka (NITK) Surathkal, Mangaluru - 575025, India}
\ead{rajanikv10@gmail.com}

\vspace{10pt}
\begin{indented}
\item[]April 2019
\end{indented}

\begin{abstract}
We investigate the thermodynamic phase transitions and heat engine efficiency in regular Bardeen AdS black hole. Interpreting cosmological constant as thermodynamic pressure, we study the thermodynamics using $T-S$ and $P-v$ plots. Specific heat studies also carried out in detail. A first order phase transition in evident from these studies. These are followed by the construction of a heat engine considering the black hole as working substance. The efficiency is obtained via a thermodynamic cycle in the $P-	V$ plane which receives and ejects heat. The heat engine efficiency is improved by adding a quintessence field. The analytical  expression for heat engine efficiency is derived in terms of quintessence dark energy parameter. This result may deepen our understanding about thermodynamics of asymptotically AdS black holes.

\end{abstract}

\vspace{2pc}
\noindent{\it Keywords}: Holographic heat engine, Regular-Bardeen AdS black hole.

%Uncomment for Submitted to journal title message
%\maketitle

\section{Introduction}
\label{sec:intro}
Black hole thermodynamics allows us to connect the quantum aspects of spacetime geometry with classical thermodynamic theory. The roots of this study trace back  to the seminal work of Hawking and Bekenstein, where the temperature and entropy for a black hole were first defined \cite{Bekenstein1972,Bekenstein1973,Bekenstein1974,hawking1975}. It is natural to think that a system which possesses thermodynamic variables such as temperature and entropy should also satisfy the law of thermodynamics. Insisting that this should be true for black holes, Bardeen, Carter and Hawking obtained the four laws of blackhole thermodynamics \cite{bardeen1973four}. A thermodynamic system can also exhibit various phases with different physical properties and phase transitions at certain critical values of the thermodynamic variables. This also turns out to be true for black holes as was first discovered by Hawking and Page in 1983 \cite{hawking1983thermodynamics}. Interestingly enough this behavior is not observed for black holes in Minkowski spacetime, but instead for black holes in an asymptotically anti-deSitter (AdS) geometry. The reason for this is that one cannot define the microcanonical ensemble for a spacetime without a boundary.

Anti-deSitter spacetime entered the spotlight when Maldacena discovered the correspondence \cite{Maldacena1999} between the \emph{classical} bulk geometry of an AdS spacetime and a \emph{quantum} conformal field theory living on the boundary of AdS. This ``AdS/CFT'' correspondence provided the first concrete construction of a quantum theory of gravity, or at the very least, a mapping between a quantum theory and a gravitational theory. Following Maldacena's discovery, the Hawking-Page transition - a first order phase transition between an AdS-Schwarzschild black hole and an AdS spacetime containing only thermal radiation - was quickly understood to be the gravitational counterpart of the confinement-deconfinement phase transition in QCD \cite{Witten:1998zw}.

Another milestone in black hole physics was the observation of a phase transition \cite{Chamblin1999,Chamblinb1999,Caldarelli2000} in charged AdS black holes, similar to that seen in a van der Waals liquid-gas system. The original form of the first law of black hole thermodynamics is written by interpreting the mass of the black hole as the internal energy. For an electrically charged rotating black hole we have, 
\begin{equation}
dM=TdS+\Omega dJ+\Phi dQ
\end{equation}
where temperature $T$ and entropy $S$ are related to surface gravity $\kappa$ and area of event horizon $A$ respectively.  $\Omega$ and $J$ are the angular velocity and  the momentum associated with rotation; $Q$ and $\phi$ are the electric charge and potential. However it was noticed that the black hole mass is more like the enthalpy $H$ than internal energy $U$, which demands a thermodynamic variable in the black hole mechanics which can be interpreted as pressure\cite{Kastor2009}. The cosmological constant $\Lambda$ was the natural candidate for thermodynamic pressure in AdS black holes for which the volume of the black hole $V$ is conjugate quantity \cite {teitelboim1985cosmological,brown1988neutralization},
\begin{equation}
P=-\frac{\Lambda }{8\pi}.
\end{equation} 
The identification of cosmological constant as pressure in the first law made it consistent with Smarr's formula. The extension of first law with $PdV$ term lead to interesting developments in black hole thermodynamics. In particular the thermodynamics of asymptotically AdS Reissner-Nordstrom black hole in the extended phase space exhibited rich phase structure analogous to van-der Waals (VdW) liquid-gas systems \cite{Dolan2011a,Dolan2011b,kubizvnak2012p}. Since then, several studies were carried out in different AdS black holes in the extended phase space and the aforementioned similarity to VdW system is observed \cite{Gunasekaran2012,BelhajChabab2012,Altamiranokubi2013,Zhao2013,Hendi2013,SChen2013}. These first order phase transitions obey Maxwell's equal area law and Clausius-Clapeyron equations \cite{SPALLUCCI2013,Belhaj2015,Li2017,Zhang14,Zhao_2015,Li2017}.

The recent developments in black hole thermodynamics are the Joule-Thomson (JT) expansion  \cite{Okcu2017,Rizwan18} and holographic heat engine in charged AdS black holes \cite{johnson2014holographic,chakraborty2018benchmarking}.
The holographic heat engines are basically traditional heat engines, but termed so by Clifford V. Johnson because their operation can be described by conformal field theories on the boundary\cite{johnson2016exact}. Through  this engine a sufficient amount of mechanical work can be drawn from heat energy in AdS black holes. It is different from  Penrose process in the rotating black holes, in which one can extract the rotational energy of the black hole from the ergosphere \cite{penrose1969,piran1975}. In the first holographic heat engine constructed by Johnson for charged AdS black hole, He calculated the efficiency of the conversion. The idea of holographic heat engine is realised in various other contexts, static dyonic and dynamic black hole \cite{jafarzade2017thermodynamic}, polytropic black hole \cite{Setare2015}, Born-Infield black holes \cite{Johnson2016}, $f(r)$ black holes \cite{Zhang2016}, Gauss-Bonnet black holes \cite{JohnsonGB2016},  higher dimensional theories \cite{Belhaj2015,Weicharged}, massive black hole \cite{Mo2018,HENDI201840}, in conformal gravity \cite{XuHaoConformal}, in 3 dimensional charged BTZ black hole \cite{Mo2017}, in rotating black holes \cite{Hennigar2017} and accelerating AdS black holes \cite{Zhang2018}.

\begin{figure*}
\begin{minipage}[b]{.5\linewidth}
\includegraphics[width=0.95\textwidth]{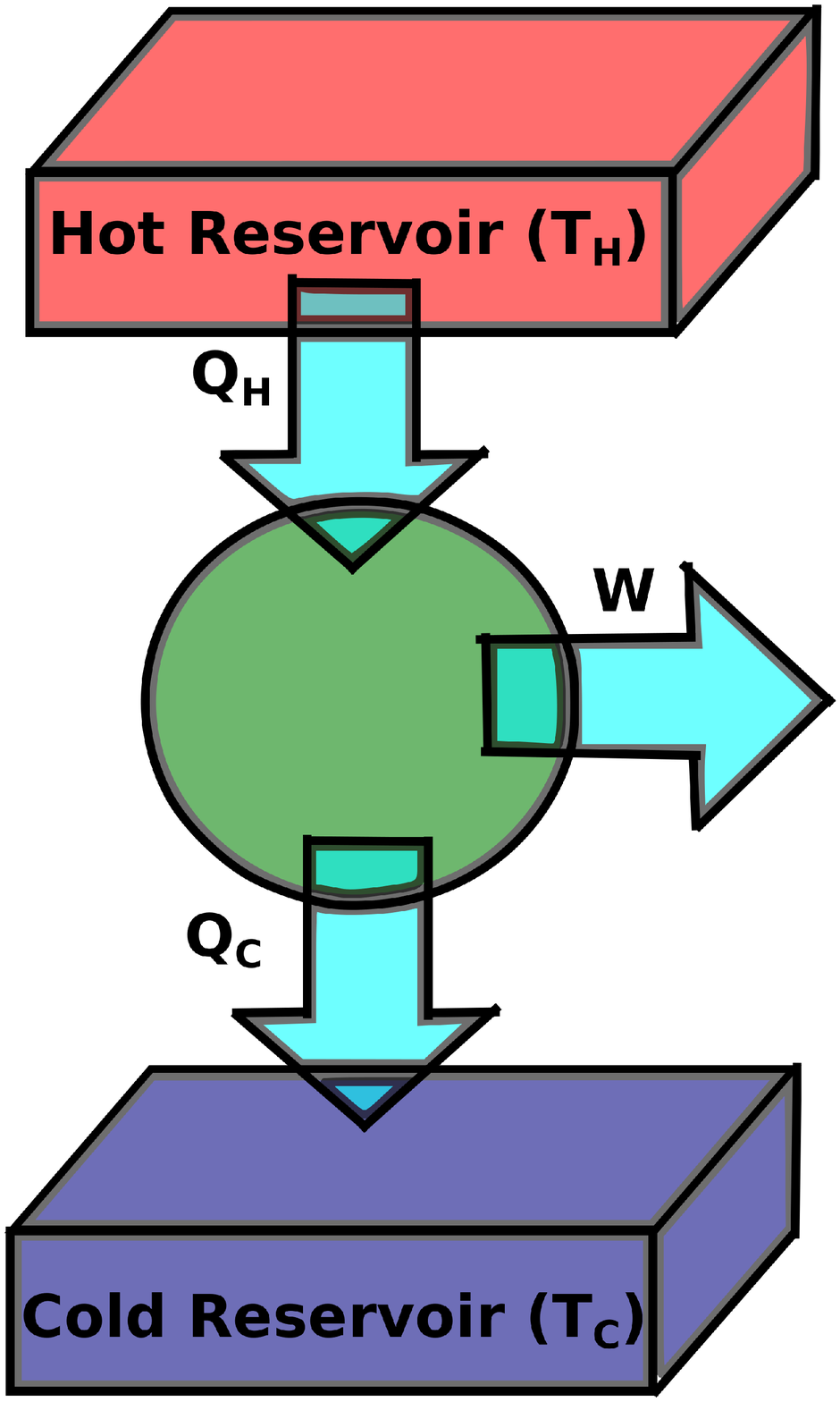}
\label{fig:heat_engine}
\end{minipage}%
\begin{minipage}[b]{.5\linewidth}
\includegraphics[width=0.95\textwidth]{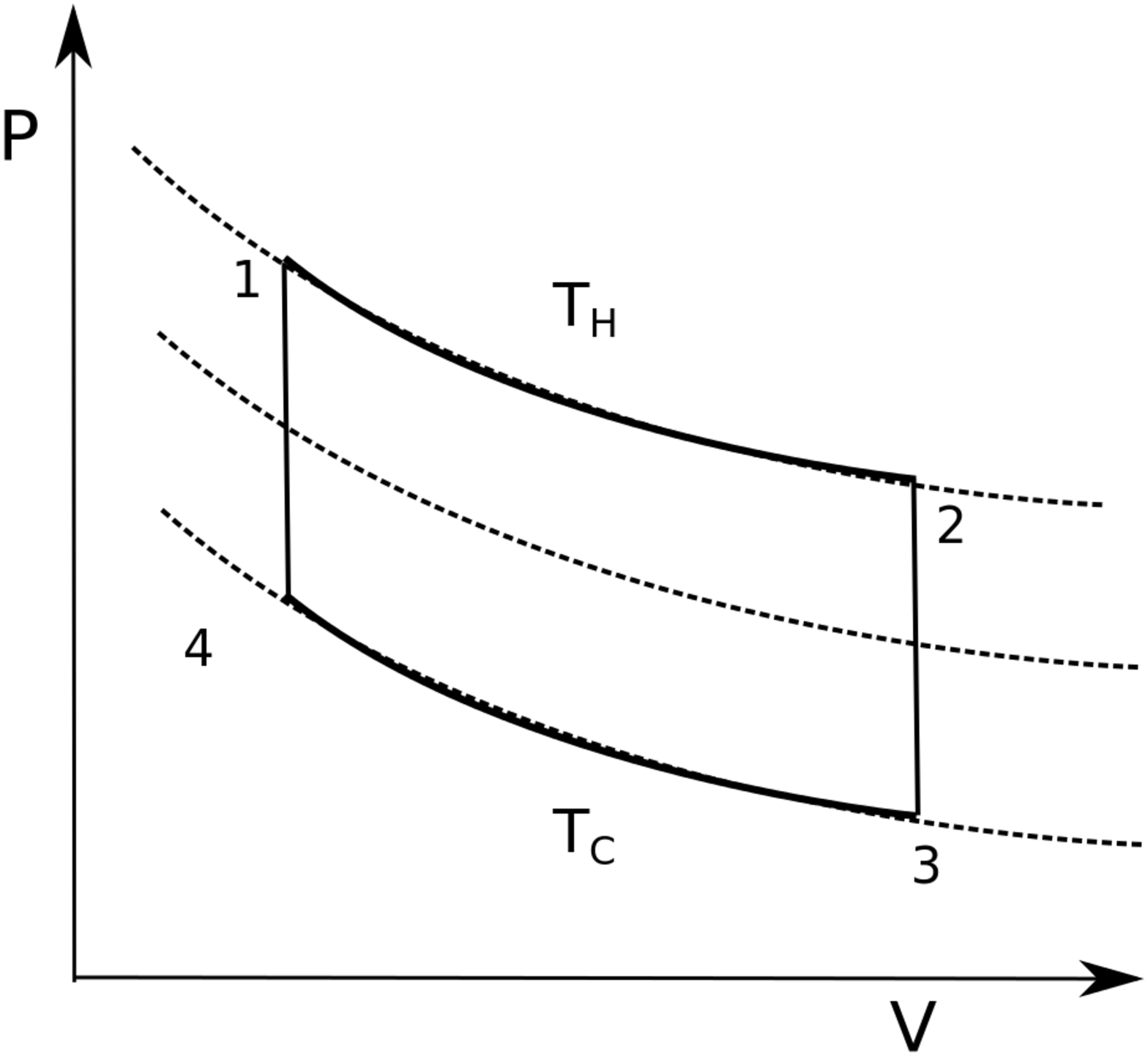}
\label{fig:carnot}
\end{minipage}
 \caption{The left figure is  the schematic diagram of heat engine and right figure is our Carnot engine cycle}
 \end{figure*}

We would like to briefly introduce heat engines. A heat engine is constructed as a closed path in the $P-V$ plane, that absorb $Q_H$ amount of heat, and exhaust $Q_C$ amount of heat (figure \ref{fig:heat_engine}). From the first law, the total mechanical work is $W=Q_H-Q_C$.  The efficiency of heat engine is given by
\begin{equation}
\eta=\frac{W}{Q_H}
\end{equation}
The maximum possible efficiency for a heat engine is estimated from a Carnot engine, which is a theoretical thermodynamic cycle.  The efficiency of Carnot engine is 
\begin{equation}
\eta_c=1-\frac{Q_C}{Q_H}=1-\frac{T_C}{T_H}  
\label{carnot eta}
\end{equation} 
where $T_C$ and $T_H$ are the lower and higher temperatures of the reservoir. This upper limit on efficiency is due to second law. 

\par Now we construct a heat engine in the context of static blackhole which has a simple heat cycle (figure \ref{fig:carnot}) with a pair of isotherms at high temperature $T_H$ and low temperature $T_C$. During isothermal expansion, $Q_H$ amount of heat is being absorbed, and it will exhaust $Q_C$ amount of heat during isothermal compression. We can connect these two temperatures by either isochoric paths as in Stirling cycle or adiabatic paths as in Carnot cycle, which is reversible. Entropy and volume for a static black hole are dependent on each other. So adiabats and isochores are alike, implies Carnot engine and Stirling engine are identical.
\par In figure (\ref{fig:carnot}) along the isotherm $1\to 2$, the amount of heat absorbed
\begin{equation}
Q_H=T_H\,\Delta S_{1\to2}=T_H\left(\frac{3}{4\pi}\right)^{\frac{2}{3}}\pi \left(V_2^\frac{2}{3}-V_1^\frac{2}{3}\right). \label{Source}
\end{equation}

Along the isotherm $3\to 4$, the amount of heat rejected
 \begin{equation}
 Q_C=T_C\,\Delta S_{3\to4}=T_C\left(\frac{3}{4\pi}\right)^{\frac{2}{3}}\pi \left(V_3^\frac{2}{3}-V_4^\frac{2}{3}\right).\label{Sink}
 \end{equation}
We choose isochores to connect those isotherms, i.e., $V_2=V_3$ and $ V_1=V_4$. Then equations (\ref{Source}) and (\ref{Sink}) leads to
 \begin{equation}
\eta=1-\frac{Q_C}{Q_H}=1-\frac{T_C}{T_H} 
 \end{equation}
which is same as the efficiency of the Carnot engine \cite{liu2017effects}.

In this paper, we extend Johnson's work  to regular Bardeen AdS black hole. 
A regular black hole has no singularity at the origin and possess an event horizon, the first of this kind was constructed by Bardeen in 1968 \cite{bardeen1968non}. The charged version of Bardeen black hole is constructed by Ayon-Beato and Garcia as magnetic solutions to Einsteins equation coupled to non linear electrodynamics \cite{AyonBeato:1998ub}. A more correct version charge free Bardeen solution was found by \cite{Hayward} in 2006. Among the several studies on the physics of regular  black holes, the thermodynamics of Regular Bardeen  black hole is investigated in \cite{MAkbar2012,2018Rizwan,TZIKAS2019219}.

This paper is organized as follows.  In section \ref{sec:thermo} we discuss the thermodynamics of regular Bardeen black hole, which is followed by the heat engine model in section \ref{sec:Heat_Engine}. In section \ref{sec:quint} the effect of quintessence on the efficiency is studied. The paper ends with results and discussions which is presented in section \ref{conclusions}.
  
\section{Thermodynamics of Regular Bardeen black hole}\label{sec:thermo}
 The regular Bardeen AdS black hole metric has the form    \cite{AYONBEATO2000149,2018Rizwan,TZIKAS2019219}
\begin{equation}
ds^2 = -f(r)dt^2+ \frac{dr^2}{f(r)} +r^2d\theta^2 +r^2 \sin^2 \theta d\phi^2
\end{equation}
where $f(r)=1-\frac{2 \mathcal{M}(r)}{r}-\frac{\Lambda r^2}{3}$ and 
 $\mathcal{M}(r)= \frac{Mr^3}{(r^2+\beta ^2)^{3/2}}$.
 $\beta$ is the monopole charge,  $M$ is the mass of the black hole, $\Lambda$ is the cosmological constant given by $-\frac{3}{\ell^2}$.

\begin{figure}
\begin{minipage}[b]{.5\linewidth}
\includegraphics[width=0.95\textwidth]{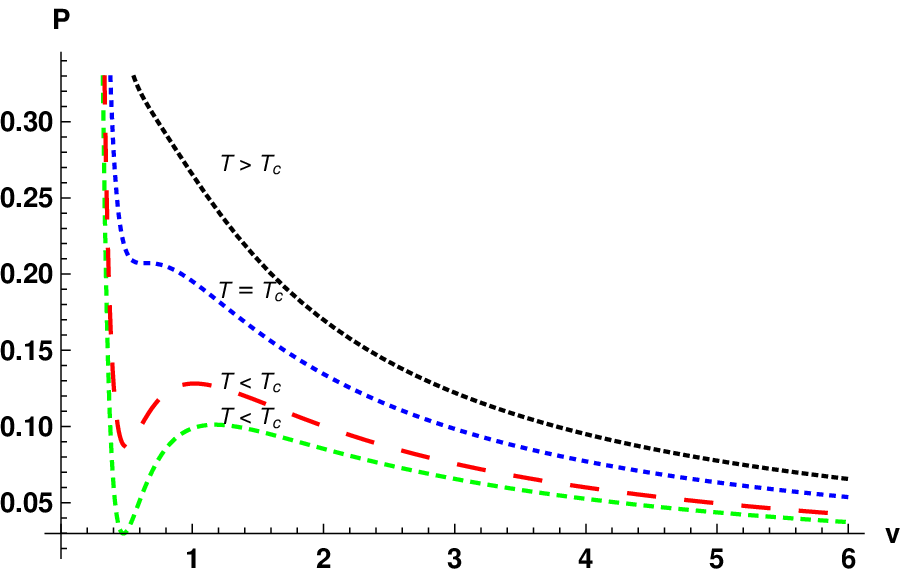}
\label{fig:PV}
\end{minipage}%
\begin{minipage}[b]{.5\linewidth}
\includegraphics[width=0.95\textwidth]{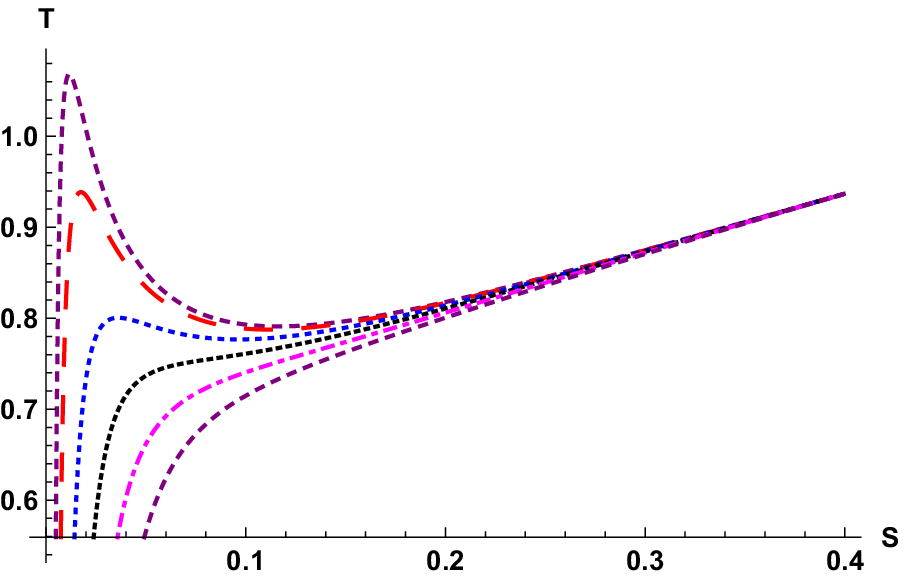}
\label{fig:TS}
\end{minipage}
\caption{(2a) P-v diagram for regular AdS black hole with $\beta=0.1$ for different values of temperature. In figure (2b) T-S plot for different values of $\beta$ is shown.}\label{PVTS}
\end{figure}

The event horizon of the Black hole is articulated by $f(r)|_{r=r_h}=0$, which gives the black hole mass

\begin{equation}
M=-\frac{\left(\beta^2+r_h^2\right)^{3/2} \left(-8 \pi  P r_h^2-3\right)}{6 r_h^2}.
\end{equation}
We can write mass in terms of entropy by using the relation $S=\pi r_h^2$
\begin{equation}
M=\frac{\left(\pi  \beta^2+S\right)^{3/2} (8 P S+3)}{6 \sqrt{\pi } S}.
\end{equation}
From the first law, we can calculate the temperature
\begin{eqnarray}
T&=\left(\frac{\partial M}{\partial S}\right)_{P,J,Q}=\frac{\sqrt{\beta^2+\frac{S}{\pi }} \left(-2 \pi  \beta^2+8 P S^2+S\right)}{4 S^2}.
\label{tseqn}
\end{eqnarray}
Rearranging the above expression and making use of the relation for apecific volume, $v=2r_h$, we obtain the equation of state
\begin{equation}
P=\frac{T}{\sqrt{4 \beta^2+v^2}}+\frac{4 \beta^2}{\pi  v^4}-\frac{1}{2 \pi  v^2}.
\label{eqnofstate}
\end{equation}
We plot the $P-v$ isotherm and $T-S$ curves using the equations (\ref{tseqn}) and  (\ref{eqnofstate})  as shown in figure (\ref{PVTS}). The $P-v$ diagram resembles the behavior of van der Waals gas. The isotherm corresponding to $T=T_C$, which is called critical isotherm, has an inflection point. The corresponding pressure and volume at that point are called critical pressure ($P_C$) and crtical volume ($v_C$), respectively. The isotherms above critical isotherm (for $T>T_C$) gradually approaces equilateral hyperbolas, which  corresponds to ideal gas case. On the other hand the lower set of isotherms (for $T<T_C$) have the positive slope region ($\partial P/\partial v >0$) which is thermodynamically unstable. The critical behavior is also apparent in $T-S$ plot. 

The critical points are obtained from the conditions, 
\begin{equation}
\frac{\partial P}{\partial v}=0 \quad, \quad \quad \frac{\partial^2 P}{\partial v^2}=0.\label{criticalform}
\end{equation}
The critical volume $v_c$, critical temperature $T_c$ and critical pressure $P_c$ are
\begin{equation}
 v_c=2 \sqrt{2} \beta \sqrt{2+\sqrt{10}},
\end{equation}
\begin{equation}
T_c=\frac{25 \left(13 \sqrt{10}+31\right)}{432 \left(2 \sqrt{10}+5\right)^{3/2} \pi  \beta},
\end{equation}
\begin{equation}
P_c=\frac{5 \sqrt{10}-13}{432 \pi  \beta^2}. 
\end{equation}
We can compute $\frac{P_c v_c}{T_c}$ ratio 
\begin{equation}
\frac{P_c v_c}{T_c}=0.381931,
\end{equation}
which almost matches with that of the van der Waals gas, which is $3/8$.
 \par One of the important physical quantity in thermodynamics is heat capacity. It is customary to define heat capacities in two different ways for a given system, i.e., at constant volume and at constant pressure. It is straight forward to show the following result,
\begin{equation}
C_V=T\left(\frac{\partial S}{\partial T}\right)_V=0 
\end{equation}
The heat capacity at constant pressure is obtained as
\begin{eqnarray}
C_P&=T\left(\frac{\partial S}{\partial T}\right)_P=\frac{2 S \left(\pi  \beta^2+S\right) \left(-2 \pi  \beta^2+8 P S^2+S\right)}{8 \pi ^2 \beta^4+4 \pi  \beta^2 S+S^2 (8 P S-1)}
\end{eqnarray}
\begin{figure}
	{
\includegraphics[width=.34\textwidth]{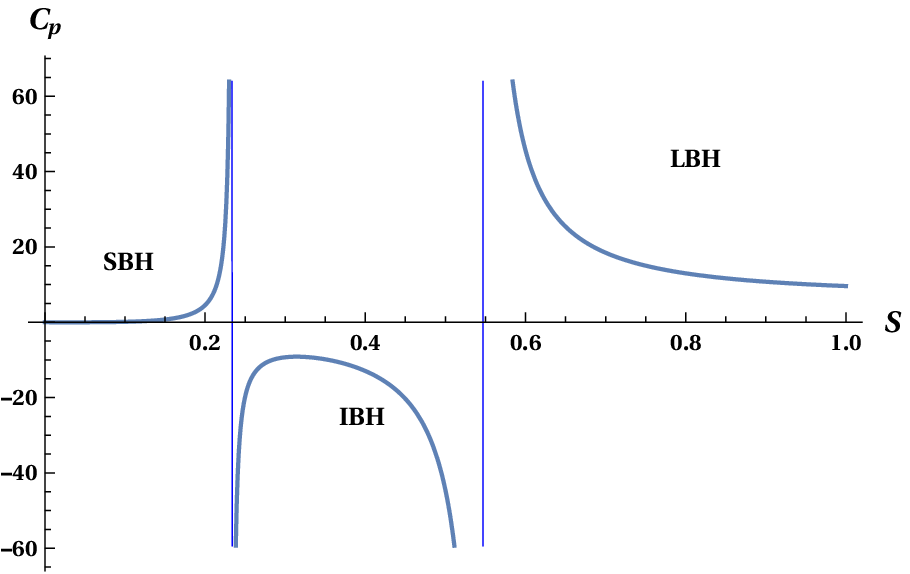}
\label{sub1}
	}
 %   \subfigure[]
    {
\includegraphics[width=.3\textwidth]{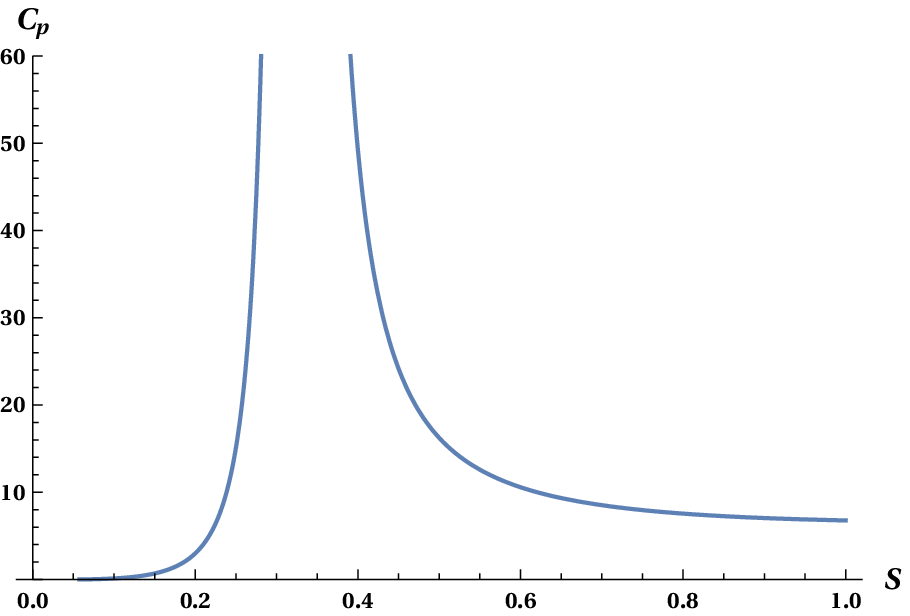}
\label{sub2}
     }
  %    \subfigure[]
         {
     \includegraphics[width=.3\textwidth]{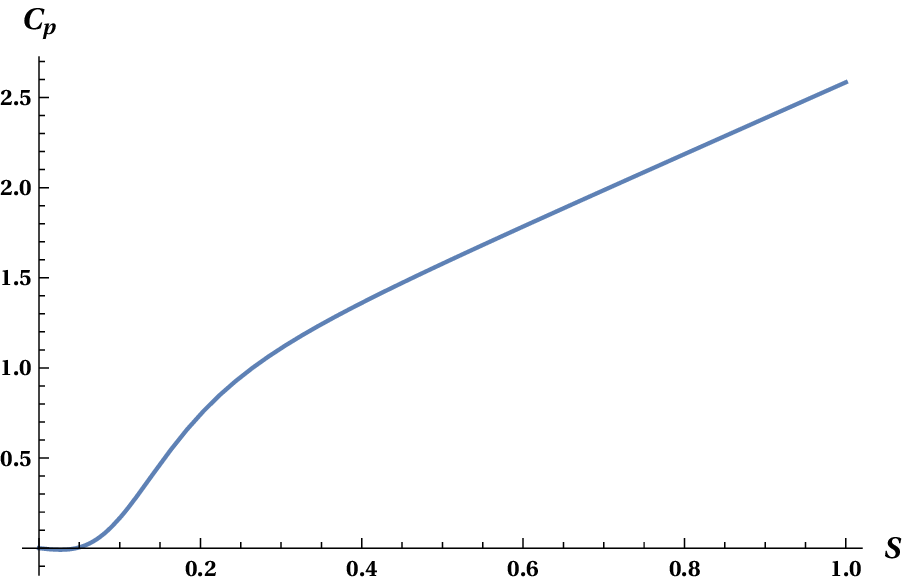}
\label{sub3}
          }
\caption{\label{CS} Specific heat versus entropy diagram for regular AdS black hole with $\beta=0.1$,  for $P<Pc$ (figure 3a), for $P=Pc$ (figure  3b) and  for $P>Pc$ (figure 3c).}
\end{figure}

The first order phase transition for the black hole is confirmed from the $C_P-S$ plot (figure \ref{CS}). The critical behavior is seen only below certain pressure $(P_c)$. There exist three distinct regions, and hence two divergence points for $P<P_C$. The positive specific heat for small black hole region (SBH) and large black hole region (LBH) means that those black holes are thermodynamically stable. Having negative specific heat, intermediate black hole region (IBH) represents unstable system. Therefore the actual phase transition takes place between small black hole and large black hole. The unstable region disapears at pressure $P=P_C$ resulting in a single divergence point.

\section{Regular black hole as a Heat Engine}
  \label{sec:Heat_Engine}
  \begin{figure}
  \begin{center}
  \includegraphics[width=6cm,height=6cm]{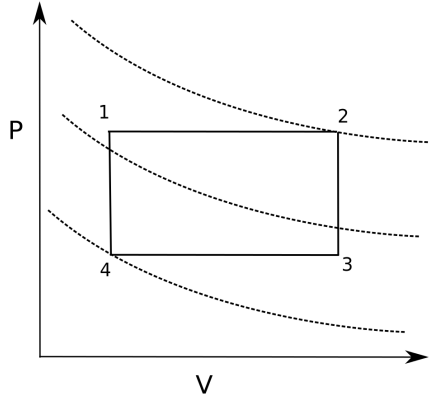}
  \caption[]{Our heat engine}
  \label{fig:Our_engine}
  \end{center}
  \end{figure}
\begin{figure}
\begin{minipage}[b]{.5\linewidth}
\includegraphics[width=0.95\textwidth]{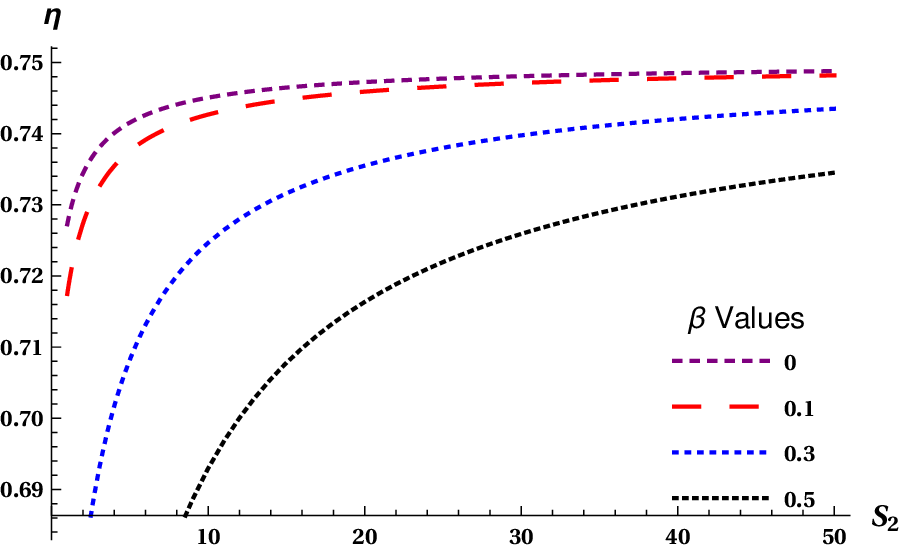}
\label{effS2}
\end{minipage}%
\begin{minipage}[b]{.5\linewidth}
\includegraphics[width=0.95\textwidth]{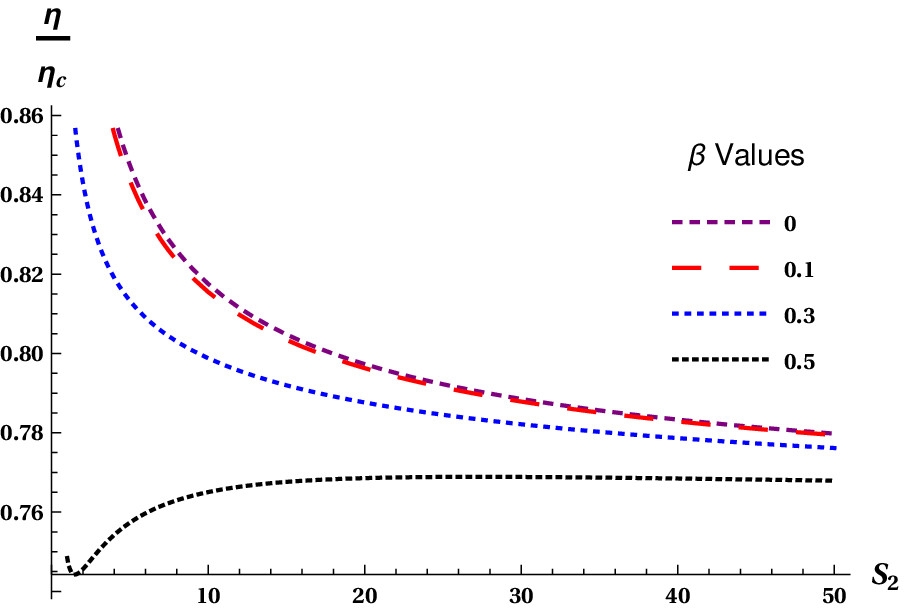}
\label{effratS2}
\end{minipage}
\caption{(5a) Efficiency versus entropy  (5b) Ratio between efficiency and Carnot efficiency ($\frac{\eta}{\eta_c}$) versus entropy diagram for regular AdS black hole with different values of $\beta$. Here we take $P_1=4,P_4=1$  and $S_1=1$.}\label{efficiencyS2}
\end{figure}
\begin{figure}
\begin{minipage}[b]{.5\linewidth}
\includegraphics[width=0.95\textwidth]{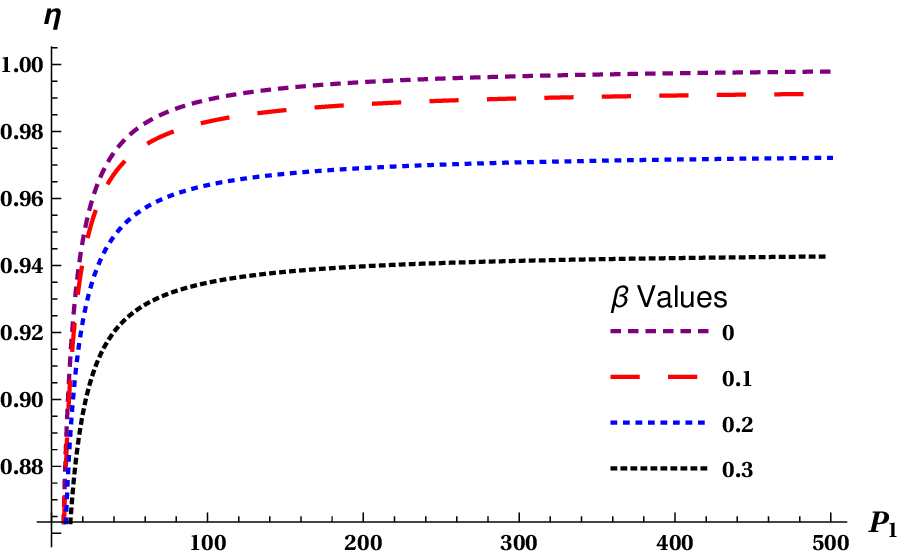}
\label{effP1}
\end{minipage}%
\begin{minipage}[b]{.5\linewidth}
\includegraphics[width=0.95\textwidth]{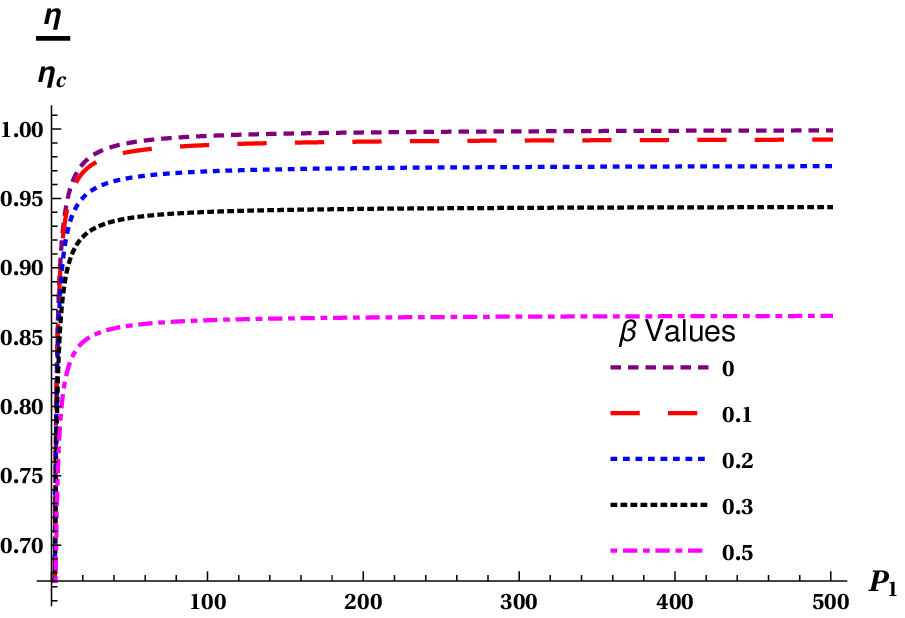}
\label{effratP1}
\end{minipage}
\caption{(6a) Efficiency versus pressure  (6b) Ratio between efficiency and Carnot efficiency ($\frac{\eta}{\eta_c}$)  versus pressure diagram for regular AdS black hole with different values of $\beta$. Here we take $P_4=1$,$S_2=4$  and $S_1=1$.}\label{efficiencyP1}
\end{figure}
\begin{figure}
\begin{minipage}[b]{.5\linewidth}
\includegraphics[width=0.95\textwidth]{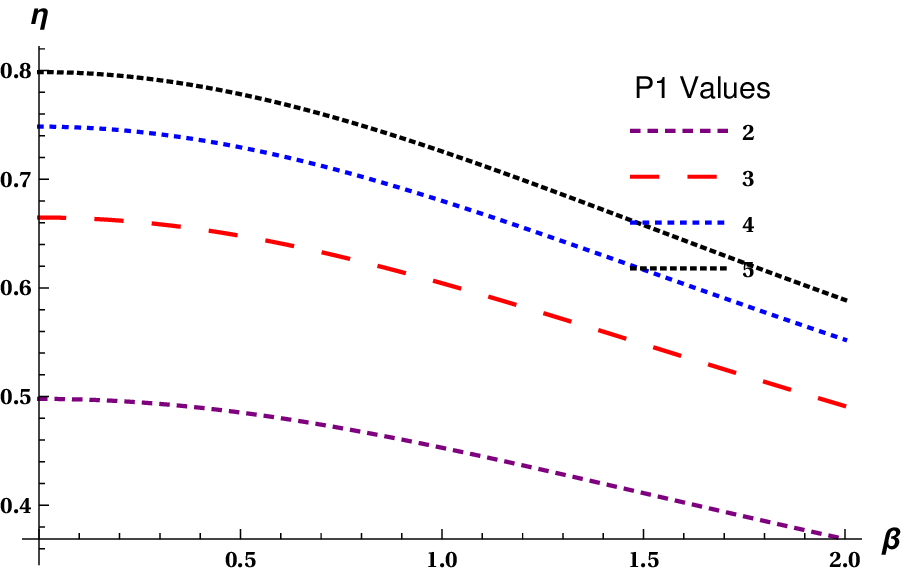}
\label{effb}
\end{minipage}%
\begin{minipage}[b]{.5\linewidth}
\includegraphics[width=0.95\textwidth]{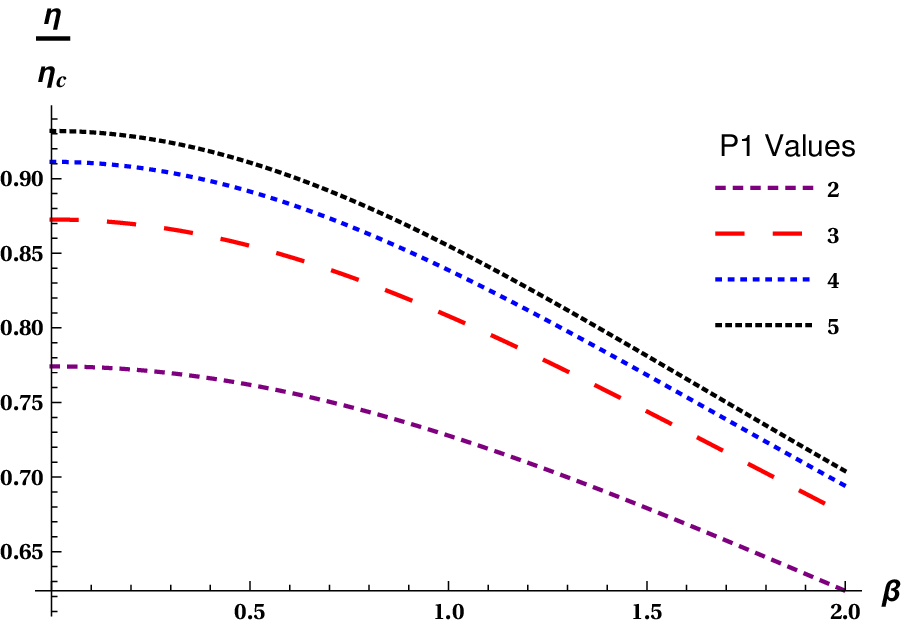}
\label{effratb}
\end{minipage}
\caption{(7a) Efficiency versus $\beta$  (7b)  Ratio between efficiency and Carnot efficiency ($\frac{\eta}{\eta_c}$) versus $\beta$ diagram for regular AdS black hole with different values of $P_1$. Here we take $S_2=20,P_4=1$  and $S_1=10$.}\label{efficiencyb}
\end{figure}

In this section, we calculate the efficiency of regular Bardeen  AdS black hole as a heat engine. As shown in figure (\ref{fig:Our_engine}), the new engine consists two isobars and two isochores/adiabats. For simplicity, consider a rectangular cycle ($1\to 2\to 3 \to 4 \to 1$). The area of the rectangle gives the work done, which reads
\begin{equation}
W=\oint PdV.
\end{equation}
The total workdone during one complete cycle 
\begin{eqnarray}
W_{tot}&=W_{1\to 2}+W_{3\to 4}=P_1(V_2-V_1)+P_4(V_4-V_3)\\
&=\frac{4}{3\sqrt{\pi}}(P_1-P_4)(S_2^{\frac{3}{2}}-S_1^{\frac{3}{2}}).
\end{eqnarray}
As the heat capacity $C_V=0$ one can conclude that no heat exchange take place in the isochoric process. Therefore we calculate only heat absorbed $Q_H$ during the process $1\to 2$ 
\begin{eqnarray}
Q_H&=\int_{T_1}^{T_2} C_P(P_1,T)dT=\int_{r_1}^{r_2} C_P(P_1,T)\frac{\partial T}{\partial r}dr\\
&=\frac{\pi  \left(\beta^2+\frac{{S_2}}{\pi }\right)^{3/2} (8 {P_1} {S_2}+3)}{6 {S_2}}-\frac{\pi  \left(\beta^2+\frac{{S_1}}{\pi }\right)^{3/2} (8 {P_1} {S_1}+3)}{6 {S_1}}.
\end{eqnarray}
The efficiency of the engine is 
\begin{eqnarray}
\eta&=\frac{W}{Q_H}
=\frac{4 ({P_1}-{P_4}) \left({S_2}^{\frac{3}{2}}-{S_1}^{\frac{3}{2}}\right)}{3 \sqrt{\pi } \left[\frac{\pi  \left(\beta^2+\frac{{S_2}}{\pi }\right)^{3/2} (8 {P_1} {S_2}+3)}{6 {S_2}}-\frac{\pi  \left(\beta^2+\frac{{S_1}}{\pi }\right)^{3/2} (8 {P_1} {S_1}+3)}{6 {S_1}}\right]}. 
\label{engine eta}
\end{eqnarray}
The efficiency of the engine $\eta$ can be compared with the maximum possible efficiency, i.e. that of Carnot engine $\eta _C$.  We take the higher temperature $T_H$ as $T_2$ and lower temperature $T_C$ as $T_4$ in equation (\ref{carnot eta}).  Then efficiency of Carnot engine is 
\begin{equation}
\eta_c=1-\frac{{T_4(P_4,S_1)}}{{T_2(P_1,S_2)}}=1-\frac{{S_2}^2 \sqrt{\pi  \beta^2+{S_1}} \left(-2 \pi  \beta^2+8 {P_4} {S_1}^2+{S_1}\right)}{{S_1}^2 \sqrt{\pi  \beta^2+{S_2}} \left(-2 \pi  \beta^2+8 {P_1} {S_2}^2+{S_2}\right)}.\label{carnot eta c}
\end{equation}
The efficiency $\eta$ versus entropy $S_2$ and the ratio $\eta /\eta _C$ versus entropy $S_2$ plots are obtained from the equations (\ref{engine eta}) and (\ref{carnot eta c}). As we can see in figure (\ref{effS2}) the heat engine efficiency monotonously increases with $S_2$ (corresponding volume $V_2$) for all values of $\beta$, which implies that the increase in volume difference between small black hole ($V_1$) and large blackhole ($V_2$) increases the efficiency. However this trend does not continue forever as the efficiency reaches saturation values after certain value of $S_2$. The dependence on $\beta$ also visible from the same figure; the rates of increment are different for different $\beta$ values. The plot $\eta /\eta _C$ versus $S_2$ is consistent with second law as it is bounded below 1. For lower values of charge $\beta$ the ratio monotonously decreases, the behavior which is reversed for higher values. We also investigate the dependence of efficiency on pressure $P_1$, the pressure at the source, which is shown in figure (\ref{efficiencyP1}). Those two figures clearly show that the efficiency of the engine will approach the maximum possible value as the pressure approaches infinity. Before concluding this section we also mention that the monopole charge $\beta$ has a negative effect on efficiency, i.e., higher the charge lower the efficiency (figure \ref{efficiencyb}).
 
\section{Efficiency of heat engine influenced by quintessence}\label{sec:quint}
Following the work of Hang Liu and Meng, we study the effect of dark energy on thermodynamics and heat engine efficiency of regular black holes \cite{liu2017effects}. Quintessence is one of the candidates for dark energy which leads to accelerated expansion of our universe \cite{Kiselev2003,Shinji2013}. The real scalar field acts as a cosmic source having equation of state $p_q=\omega \rho _q$ ($-1< \omega_q < -1/3$).
The density of quintesence field is given by 
\begin{equation}
\rho _q=-\frac{a}{2}\frac{3\omega _q}{r^{3(\omega _q+1)}}
\label{rho}
\end{equation}
Kiselev was the first to study effects of quintessence on a black hole  \cite{Kiselev2003}. Since then there were many studies in black holes surrounded by quintessence, to mention few, in the contexts of gauge gravity duality \cite{2013chen} and quasi normal modes \cite{2005chen}. Thermodynamics phase transitions in Reissner-Nordstr{\"o}m and regular  black holes with this exotic field were studied in \cite{WeiYi2011,QuienRNThomas2012,LI2014,Fan2017,Saleh2018,Rodrigue2018lzp}.
When we include quintessence term in the metric of regular Bardeen AdS black hole, $f(r)$ is modified to
\begin{equation}
f(r)=1-\frac{2 m r^2}{\left(\beta^2+r^2\right)^{3/2}}-\frac{a}{ r^{3 \omega_q+1}}-\frac{\Lambda r^2}{3}.
\end{equation}
Where $a$ is the normalization constant related to quintessence density, $\omega_q$ is the state parameter \cite{2018Rizwan}. 

Now we proeed as in the earlier sections to obtain an expression for efficiency of the engine. Using the defining condition of event horizon, $f(r_h)=0$, one can calculate the black hole mass as
\begin{equation}
M=\frac{\left(\pi  \beta^2+S\right)^{3/2} S^{-\frac{3}{2} (\omega_q+1)} \left[(8 P S+3) S^ {(3 \omega_q+1)/2}-3 a \pi ^{(3 \omega_q+1)/2}\right]}{6 \sqrt{\pi }}.
\end{equation}
This expression along with first law enables us to write the temperature,
\begin{eqnarray}
T=\frac{1}{4} \sqrt{\beta^2+\frac{S}{\pi }} S^{-\frac{3 \omega_q}{2}-\frac{5}{2}} &\left[3 a \pi ^{\frac{3 \omega_q}{2}+\frac{1}{2}} \left(\pi  \beta^2 (\omega_q+1)+S \omega_q\right)\right.\nonumber \\
&\quad\quad\quad +\left. S^{\frac{3 \omega_q}{2}+\frac{1}{2}} \left(-2 \pi  \beta^2+8 P S^2+S\right)\right].
\end{eqnarray}
The heat capacity at constant pressure is
\begin{equation}
\textstyle{C_P=\frac{2 S \left(\pi  \beta^2+S\right) \left[3 a \pi ^{\frac{3 \omega_q+1}{2}} \left(\pi  \beta^2 (\omega_q+1)+S \omega_q\right)+S^{\frac{3 \omega_q+1}{2}} \left(-2 \pi  \beta^2+8 P S^2+S\right)\right]}{S^{\frac{3 \omega_q+1}{2}} f_1(S)-3 a \pi ^{\frac{3 \omega_q+1}{2}} \left(f_2(S)\right]}}
\end{equation}
where
\begin{eqnarray}
f_1(S)=8 \pi ^2 \beta^4+4 \pi  \beta^2 S+S^2 (8 P S-1)\nonumber \\
f_2(S)=\pi ^2\beta^4 \left(3 \omega_q^2+8 \omega_q+5\right)+2 \pi  \beta^2 S \left(3 \omega_q^2+5 \omega_q+2\right)+S^2 \omega_q (3 \omega_q+2).\nonumber
\end{eqnarray}
Then we compute the heat $Q_H$ along the process $1\to 2$ (the earlier arguments on no heat transfer for isochoric processes still holds),
\begin{equation}
Q_H= \int_{T_1}^{T_2}C_P(P_1,T)dT=\int_{S_1}^{S_2}\left(\frac{\partial T}{\partial S}\right)dS=\int_{S_1}^{S_2}TdS=M_2-M_1
\end{equation}

\begin{eqnarray}
Q_H=\frac{1}{6\sqrt{\pi}}&\left\lbrace \left(\pi  \beta^2+S_1\right)^{3/2} S_1^{-\frac{3}{2}  (\omega_q+1)} \left[3 a \pi ^{\frac{3 \omega_q+1}{2}}-(8 P_1 S_1+3) S_1^{\frac{3 \omega_q+1}{2}}\right]\right. \nonumber \\
&+\left.\left(\pi \beta^2+S_2\right)^{3/2} S_2^{-\frac{3}{2}(\omega_q+1)} \left[(8 P_1 S_2+3) S_2^{\frac{3 \omega_q+1}{2}}-3 a \pi ^{\frac{3 \omega_q+1}{2}}\right]\right\rbrace \nonumber \\
\end{eqnarray}
Having all the required quantities, the heat engine efficiency is expressed interms of quintessence parameters $a$ and $\omega _q$ as
\begin{equation}
\eta =\frac{8 ({P_1}-{P_4}) \left({S_2}^{3/2}-{S_1}^{3/2}\right)}{f(S_1)-f(S_2)}
\end{equation}
where
\begin{equation*}
f(S)=\left(\pi  \beta^2+{S}\right)^{\frac{3}{2}} {S}^{-\frac{3}{2} (\omega_q+1)} \left[3 a \pi ^{\frac{3 \omega_q}{2}+\frac{1}{2}}-(8 {P_1} {S}+3) {S}^{\frac{3 \omega_q}{2}+\frac{1}{2}}\right].
\end{equation*}
The efficiency of the Carnot engine is also obtained as earlier,
\begin{equation}
\eta_C=1-\frac{g(S_1,P_4)}{g(S_2, P_1)}
\end{equation}
with
\begin{eqnarray}
g(S,P)=\frac{\sqrt{\pi \beta^2+S}}{S^{\frac{3 \omega_q}{2}+\frac{5}{2}}}
&\Bigg[3 a \pi ^{\frac{3 \omega_q}{2}+\frac{1}{2}} \left(\pi \beta^2 (\omega_q+1)+S \omega_q\right)\nonumber \\
& +S^{\frac{3 \omega_q}{2}+\frac{1}{2}} \left(-2 \pi  \beta^2+8 P S^2+S\right)\Bigg].
\end{eqnarray}

\begin{figure}
\begin{minipage}[b]{.5\linewidth}
\centering
\includegraphics[width=0.95\textwidth]{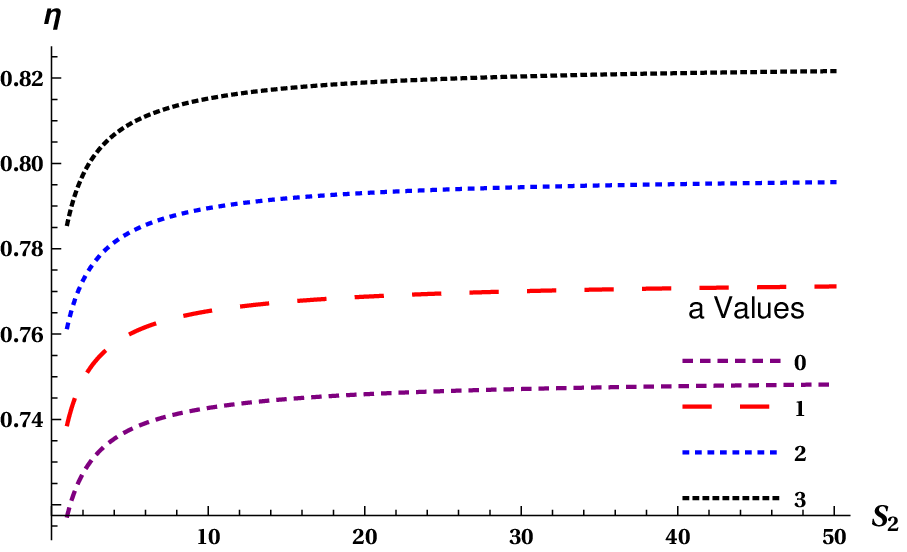}
\label{effS2a}
\end{minipage}%
\begin{minipage}[b]{.5\linewidth}
\centering
\includegraphics[width=0.95\textwidth]{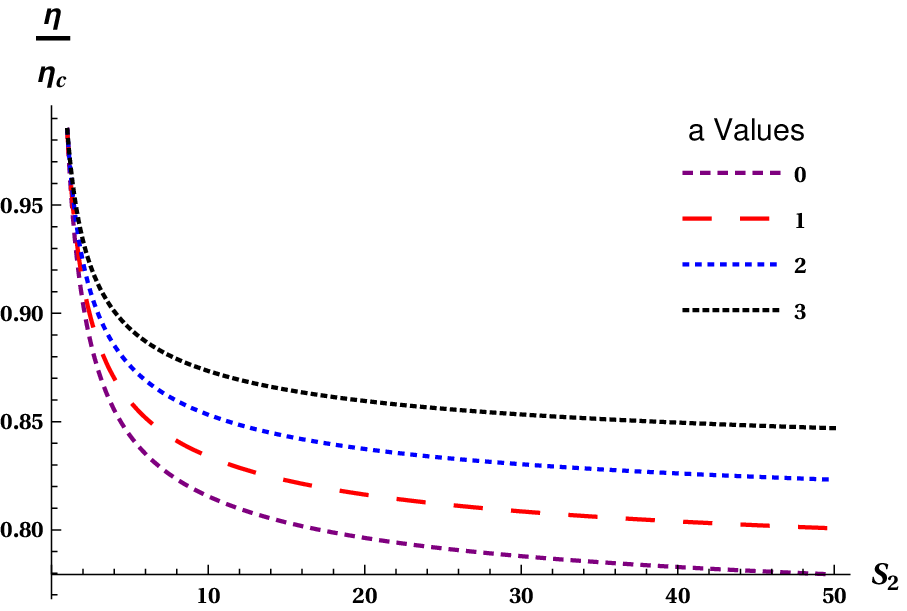}
\label{effratioS2a}
\end{minipage}
\caption{(8a) Efficiency versus entropy (8b) Ratio between efficiency and Carnot efficiency ($\frac{\eta}{\eta_c}$)  versus entropy diagram for regular AdS black hole with different values of $a$. Here we take $P_1=4,P_4=1$  $S_1=1$, $\omega_q=-1$ and $\beta=0.1$.}\label{efficiencyS2a}
\end{figure}

\begin{figure}
\begin{minipage}[b]{.5\linewidth}
\centering
\includegraphics[width=0.95\textwidth]{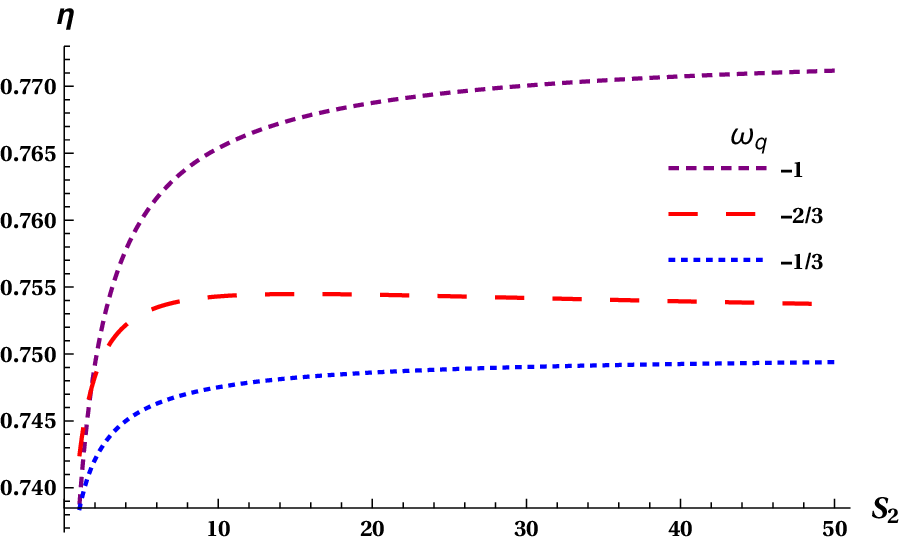}
\label{EffS2w}
\end{minipage}%
\begin{minipage}[b]{.5\linewidth}
\centering
\includegraphics[width=0.95\textwidth]{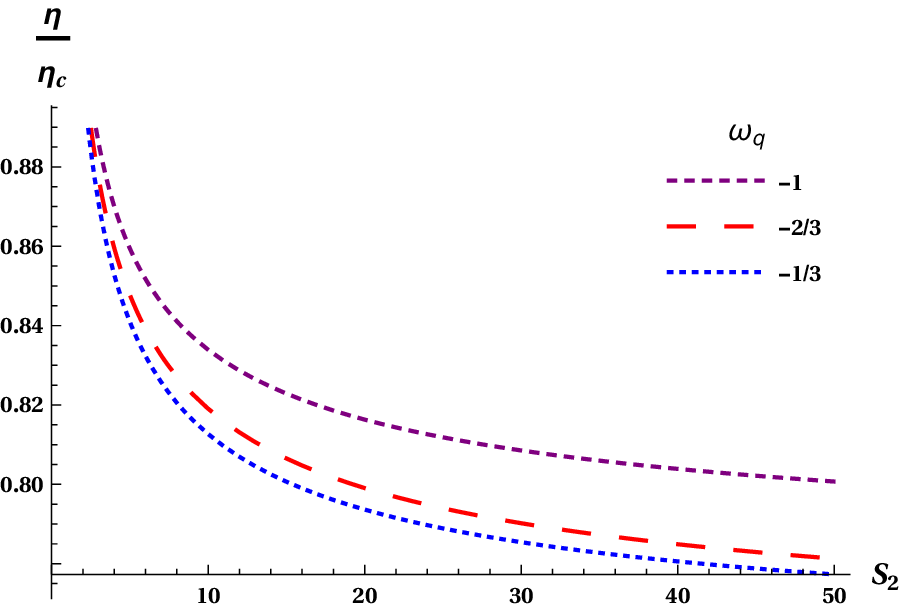}
\label{EffratioS2w}
\end{minipage}
\caption{(9a) Efficiency versus entropy (9b) Ratio between efficiency and Carnot efficiency ($\frac{\eta}{\eta_c}$)  versus entropy diagram for regular AdS black hole with different values of $\omega_q$. Here we take $P_1=4,P_4=1$  $S_1=1$, $a=1$ and $\beta=0.1$.}\label{efficiencyS2w}
\end{figure}

\begin{figure}
\begin{minipage}[b]{.5\linewidth}
\centering
\includegraphics[width=0.95\textwidth]{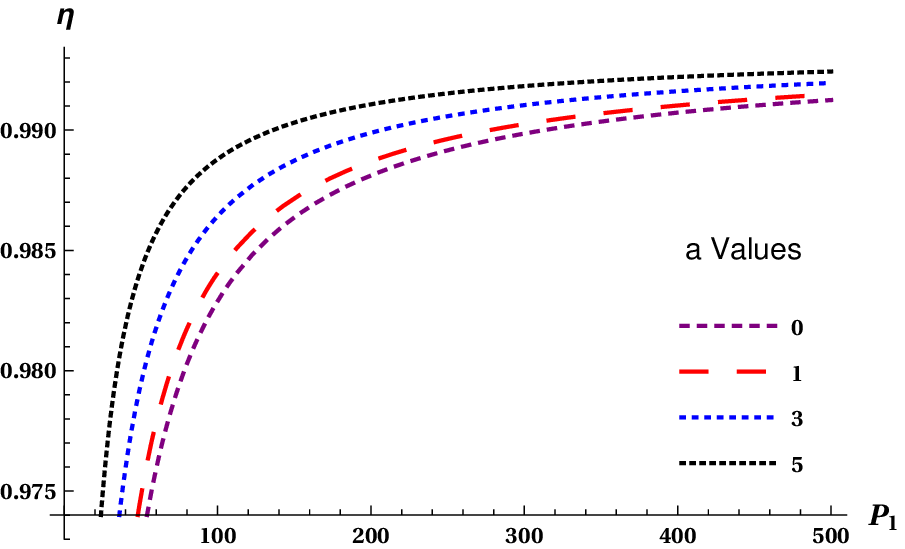}
\label{effP1a}
\end{minipage}%
\begin{minipage}[b]{.5\linewidth}
\centering
\includegraphics[width=0.95\textwidth]{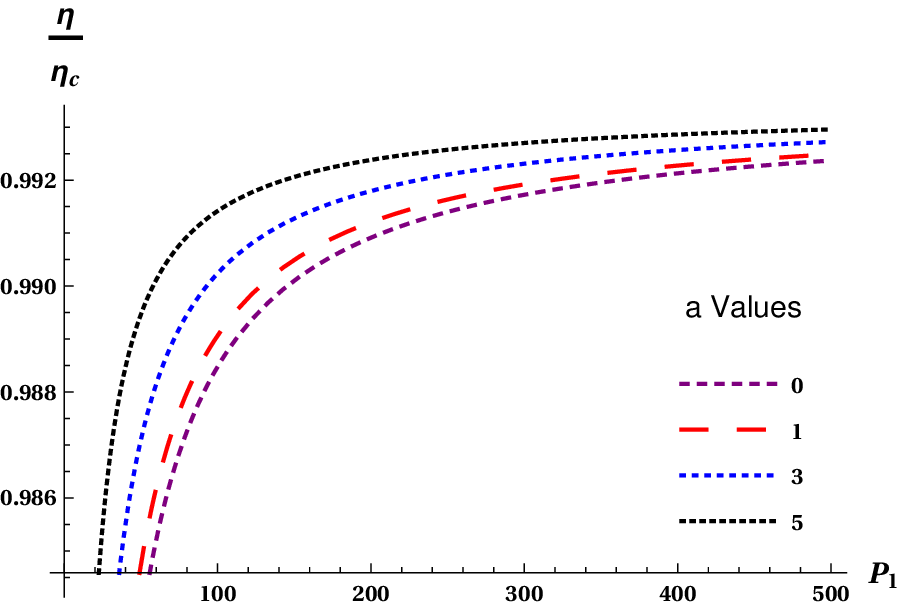}
\label{effratioP1a}
\end{minipage}
\caption{(10a) Efficiency versus pressure (10b) Ratio between efficiency and Carnot efficiency ($\frac{\eta}{\eta_c}$)  versus pressure diagram for regular AdS black hole with different values of $a$. Here we take $P_4=1$  $S_1=1$, $S_2=4$, $\omega_q=-1$ and $\beta=0.1$.}\label{efficiencyP1a}
\end{figure}

\begin{figure}
\begin{minipage}[b]{.5\linewidth}
\centering
\includegraphics[width=0.95\textwidth]{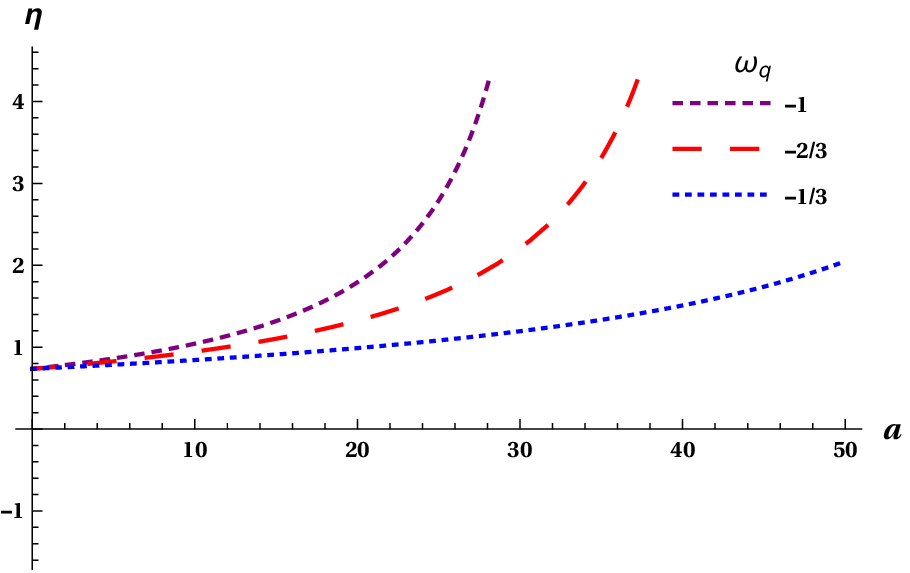}
\label{effaw}
\end{minipage}%
\begin{minipage}[b]{.5\linewidth}
\centering
\includegraphics[width=0.95\textwidth]{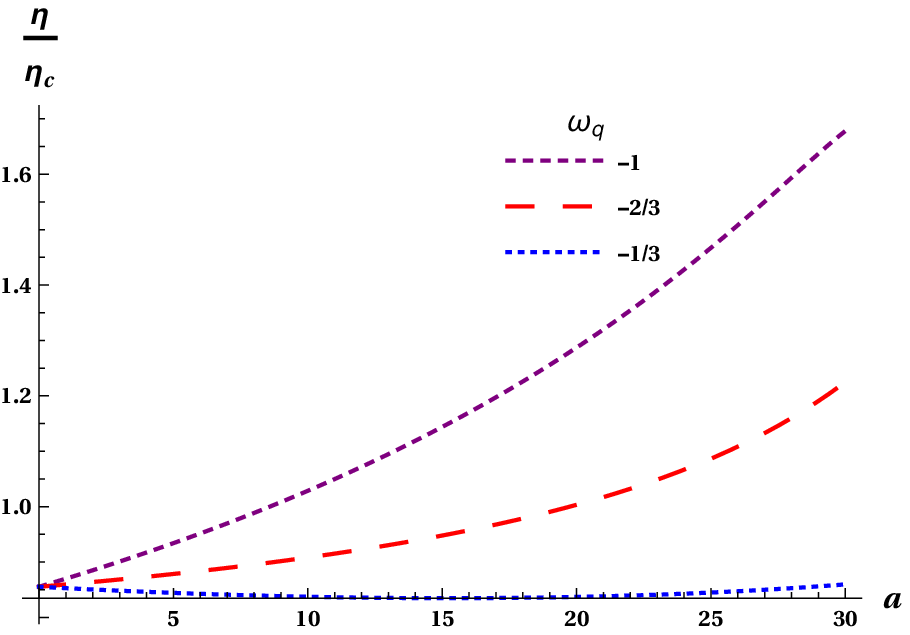}
\label{effratioaw}
\end{minipage}
\caption{(11a) Efficiency versus normalization factor $a$ (11b) Ratio between efficiency and Carnot efficiency ($\frac{\eta}{\eta_c}$)  versus normalization factor $a$ diagram for regular AdS black hole with different values of $\omega_q$. Here we take $P_1=4,P_4=1$  $S_1=1$, $S_2=4$ and $\beta=0.1$.}\label{efficiencyaw}
\end{figure}

\begin{figure}
\begin{minipage}[b]{.5\linewidth}
\centering
\includegraphics[width=0.95\textwidth]{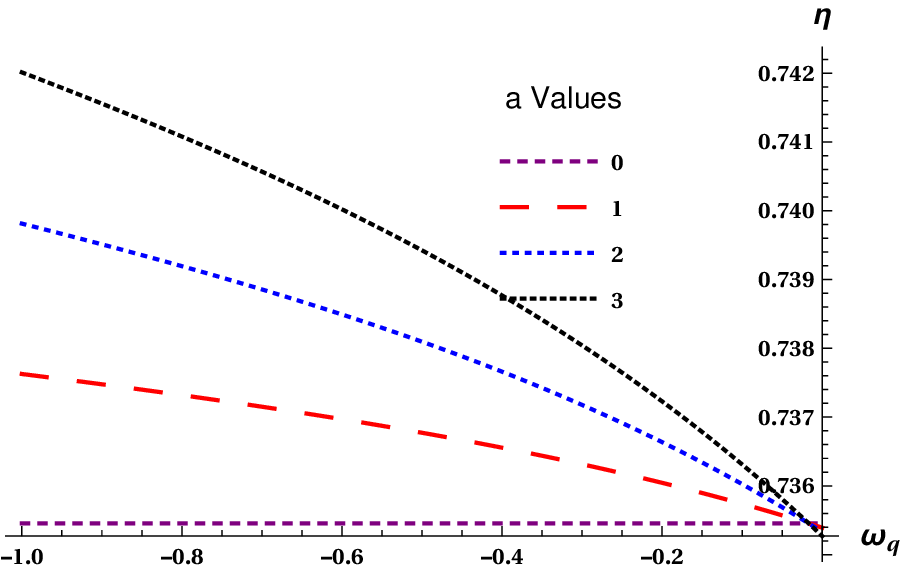}
\label{effwa}
\end{minipage}%
\begin{minipage}[b]{.5\linewidth}
\centering
\includegraphics[width=0.95\textwidth]{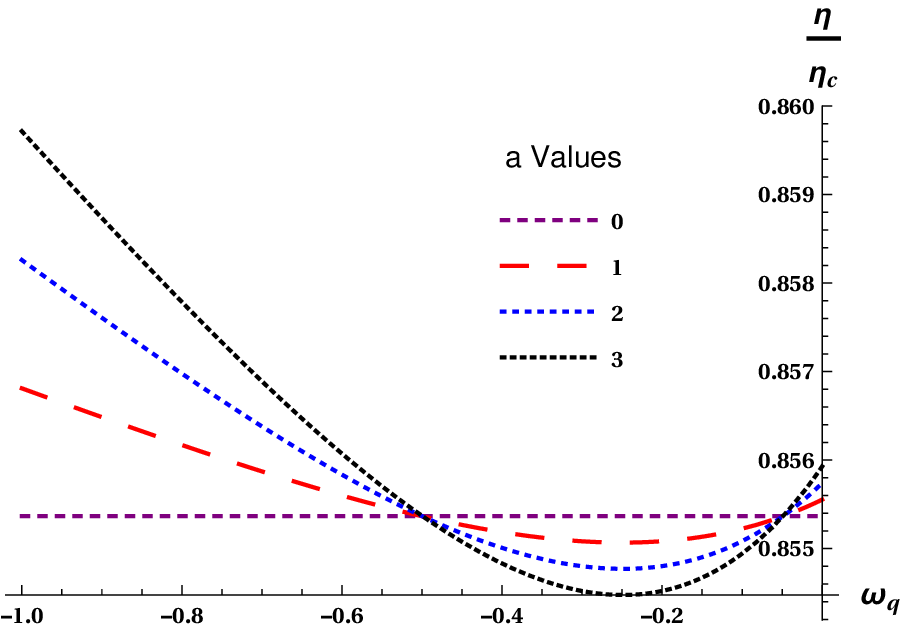}
\label{effratiowa}
\end{minipage}
\caption{(12a) Efficiency versus state parameter $\omega$ (12b) Ratio between efficiency and Carnot efficiency ($\frac{\eta}{\eta_c}$)  versus state parameter $\omega_q$ diagram for regular AdS black hole with different values of $a$. Here we take $P_1=4,P_4=1$  $S_1=1$, $S_2=4$ and $\beta=0.1$.}\label{efficiencywa}
\end{figure}

The heat engine efficiency depends on pressure, normalization constant, entropy, state parameter and monopole charge. The above expressions reduce to the previous case when $a=0$ and $\omega _q=0$. There is a significant increment in the efficiency against $S_2$ when we increase the normalisation constant $a$ with other parameters being fixed (figure \ref{efficiencyS2a}). The change is visible in the ratio plot also. The plot for efficiency versus $S_2$ for different values of  $\omega _q$ show similar functional behavior (figure \ref{efficiencyS2w}). But there is a difference in the physical effect, higher values of $\omega _q$ lead to smaller efficiency. This is not surprising because the quintessence density ($\rho _q$) decreases with increasing $\omega _q$ (equation \ref{rho}). We study the role of pressure $P_1$ on $\eta$ and $\eta /\eta _C$ (figure \ref{efficiencyP1a}), where the functional appearance remains same. But the approach towards limiting value 1 is faster than that of without quintessence. 

At the final stage of our investigation we focus on the action of $a$ and $\omega _q$. 	For all three values of $\omega _q$ efficiency increases exponentially, when plotted against $a$. The scenario remains same for ratio plot with an exception at $\omega _q=-1/3$, which has a slight decaying nature intially. We note that in these two plots the efficiency shoots over unity which is a clear violation of second law of thermodynamics. To avoid this unphysical situation we must be careful enough to chose quintessence parameters. In figure (\ref{efficiencywa}) we present the effect of $\omega _q$ on $\eta$ and $\eta /\eta _C$ for different values of $a$. In the light of earlier point, quintessence density ($\rho _q$) decreases with increasing $\omega _q$, the efficiency is higher for lower smaller values of $\omega _q$. This inference is considering the physically meaningful range $-1< \omega_q < -1/3$.

\section{Conclusions and Discussions}\label{conclusions}
We study the phase transitions and heat engine efficiency for regular Bardeen AdS black hole. The main focus of this work being on black hole as a heat engine, the efficiency calculations are extended to the case with quintessence. In the extended phase space we investigate the critical behavior via the $P-v$ and $T-S$ plots. The plots resemble with that of van der Waals gas. This was followed by the calculations of specific heats at constant volume and pressure. $C_P-S$ plot affirms the first order phase transition below a critical pressure. The vanishing $C_V$ allows us to neglect the heat exchange in isochores in the study of engines later.

Heat engines are constructed by taking regular Bardeen black hole as working substance. A cycle in $P-v$ plane is assigned for the black hole with two isotherms and two isochores. The efficiency of the engine is calculated by using the work done and heat absorbed during the cycle. As it is customary to compare the efficiency of any engine with Carnot engine we have compared our results with the corresponding Carnot efficiency. Detailed analysis of the dependence of $\eta$ on $S_2$, $P_1$ and $\beta$ are done. Among the several observations we emphasize that, the increase in entropy difference between small black hole ($S_1$) and large black hole ($S_2$) increases the efficiency. Bigger pressure difference also increases the efficiency. In all these studies the efficiency remain bounded below unity which is consistent with second law. We have made a successful attempt to improve the efficiency of engine by adding a quintessence field.  

\section{Acknowledgements}
The authors R.K.V., A.R.C.L. and N.K.A. would like to thank the department of physics, National Institute of Technology Karnataka. The author N.K.A. also thank UGC, Govt. of India for financial support through SRF scheme.
\section*
{References}
\label{refs}

\bibliographystyle{iopart-num}
\bibliography{heat_engine.bib}
\end{document}